\documentstyle[12pt]{article}

\renewcommand{\baselinestretch}{1.4}

\newfont{\Mb}{msbm10}                   
\newcommand{\R}{\mbox{\Mb\symbol{82}}}

\begin{document}

\renewcommand{\baselinestretch}{1.1}

\title{\Large \bf Causal Anomalies in Kaluza-Klein 
                         Gravity Theories \\} 
\author{
M.J. Rebou\c{c}as\thanks{
{\sc internet: reboucas@cat.cbpf.br}  } \ \ 
and  \  
A.F.F Teixeira\thanks{
{\sc internet: teixeira@novell.cat.cbpf.br}  }  \\ 
\\  
Centro Brasileiro de Pesquisas F\'\i sicas\\
             \ \ Departamento de Relatividade e Part\'\i culas \\
                 Rua Dr.\ Xavier Sigaud 150 \\
                 22290-180 Rio de Janeiro -- RJ, Brazil \\ \\
       }

\date{\today}  

\maketitle

\begin{abstract}
Causal anomalies in two Kaluza-Klein gravity theories
are examined, particularly as to whether these theories 
permit solutions in which the causality principle is violated. 
It is found that similarly to general relativity the field 
equations of the space-time-mass Kaluza-Klein (STM-KK) gravity 
theory do not exclude violation of causality of G\"odel type, 
whereas the induced matter Kaluza-Klein (IM-KK) gravity rules out 
noncausal G\"odel-type models.
The induced matter version of general relativity is shown to be
an efficient therapy for causal anomalies that occurs in a wide 
class of noncausal geometries.
Perfect fluid and dust G\"odel-type solutions of the STM-KK field 
equations are studied.
It is shown that every G\"odel-type perfect fluid solution  
is isometric to the unique dust solution of the STM-KK field 
equations.
The question as to whether 5-D G\"odel-type non-causal geometries 
induce any physically acceptable 4-D energy-momentum tensor is also 
addressed. 
\end{abstract}

\newpage

\section{Introduction}   \label{intro}
\setcounter{equation}{0}

Kaluza-Klein-type theories in five or more dimensions have
been of interest in a few contexts.
In the framework of gauge theories they have been used in the 
quest for unification of the fundamental interactions in physics.
The idea that the various interactions in nature might
be unified by enlarging the dimensionality of the space-time
has a long history that goes back to the works of Kaluza and 
Klein~\cite{Kaluza1921}~--~\cite{Klein1926}.
They showed how one could unify Einstein's theory of gravitation and 
Maxwell's theory of electromagnetism in a five-dimensional framework.
The higher-dimensional Kaluza-Klein approach to unification was 
later used by the unified-field theorists, especially among those 
investigating the eleven-dimensional supergravity and ten-dimensional
superstrings.

In 1983, Wesson~\cite{Wesson83,Wesson84} has given an impetus to the 
study of Kaluza-Klein type theories by investigating a five-dimensional 
Kaluza-Klein {\em theory of gravitation\/} with a variable rest mass. 
In this theory we have five-dimensional (5-D) space-time-mass manifolds, 
and the fifth dimension is a convenient mathematical way of geometrizing 
the rest mass and  of allowing one to study the possibility that it may 
be variable. The four-dimensional (4-D) general relativity theory is 
recovered when the rate of change of the rest mass is zero. We shall 
refer to this theory as space-time-mass Kaluza-Klein gravity theory 
(STM-KK gravity theory, for short).
Ever since the foundations of the STM-KK  gravity theory were laid, 
there have been investigations on its potentialities, particularly 
as concerns to its consequences for astrophysics and  cosmology, 
the confrontation between theory and observation, its consistency 
with Mach's principle, the solutions to its field equations, and 
the like~\cite{Wesson84}~--~\cite{Fukui92}. For fair list of 
references on the STM-KK theory see Overduin and Wesson~\cite{Overduin97}.

More recently, Wesson and co-workers~\cite{Wesson90,WessonLeon92a}
have introduced a new approach to general relativity, 
conceptually in line with an old idea due to Einstein%
~\cite{Einstein56}, in which the matter 
and its role in the determination of the space-time 
geometry is given from a purely five-dimensional geometrical 
point of view. In their 5-D version of general relativity the 
field equations are given by 
\begin{equation} \label{5DfeqsG}
\widehat{G}_{AB} = 0 \;.
\end{equation}
Henceforth, the five-dimensional geometrical objects are denoted by 
overhats and Latin letters are 5-D indices and run from $0$ to $4$.
The 5-D vacuum field equations (\ref{5DfeqsG}) give 
rise to both curvature and matter in 4-D. Indeed, it can be 
shown~\cite{WessonLeon92a} that it is always possible to rewrite 
the field equations (\ref{5DfeqsG}) as a set of fifteen 
equations, ten of which are precisely Einstein's field 
equations
\begin{equation}
G_{\alpha \beta} = \kappa \,\; T_{\alpha \beta} \label{ein} 
\end{equation}
in 4-D with an {\em induced\/} 
energy-momentum 
\begin{eqnarray} \label{Tinduced}
T_{\alpha\beta} & = & \frac{\phi_{\alpha\,;\:\beta}}{\phi} - 
\frac{\varepsilon}{2\,\phi^2} 
   \left\{ \frac{\phi^{*} \, g^{*}_{\alpha\beta}}{\phi}
   - g^{**}_{\alpha\beta} + g^{\gamma\delta} \, 
   g^{*}_{\alpha\gamma} \,  g^{*}_{\beta\delta} 
-  \frac{g^{\gamma\delta} \, 
   g^{*}_{\gamma\delta} \, g^{*}_{\alpha\beta}}{2}  
   \right. \nonumber \\
              &   &
+ \; \left.
\frac{g^{}_{\alpha\beta}}{4^{} {}_{} } \, \left[\,g^{*}{}^{\gamma\delta} \, 
 g^{*}_{\gamma\delta} + (g^{\gamma\delta_{}} \,g^{*}_{\gamma\delta})^2
       \, \right]  \,\right\} \;,
\end{eqnarray} 
where from now on Greek letters denote 4-D indices and run from 
$0$ to $3$, $g_{44} \equiv \varepsilon\, \phi$ with $\varepsilon=\pm 1 $,
$\phi_\alpha \equiv \partial \phi / \partial x^\alpha$, a star
denotes $\partial / \partial x^4$, and a semicolon  denotes the usual
4-D covariant derivative. Obviously,
the remaining five equations (a wave equation and four conservation 
laws) are automatically satisfied by any solution of the 5-D 
vacuum equations~(\ref{5DfeqsG}).
Thus, not only the matter  but also its role in the 
determination of the geometry of the 4-D space-time can be considered 
to have a five-dimensional geometrical origin. This is an elegant 
step towards the realization of the Einstein's vision%
~\cite{Einstein56}~--~\cite{Salam80} of nature as {\em pure geometry\/} 
in that it unifies the gravitational field with its source (not just 
with other fields) within a purely 5-D geometrical framework.
We shall refer to this 5-D version of general relativity as 
induced matter Kaluza-Klein gravity theory (IM-KK gravity theory, for
short). 
The IM-KK theory has become a focus of a recent research field%
~\cite{Overduin97}.
The basic features of the theory have been explored by Wesson and 
others~\cite{Leon93}~--~\cite{Wesson96a}, whereas the implications 
for cosmology and astrophysics have been investigated by a number 
of researchers~\cite{Wesson92b}~--~\cite{Liu96b}. 
For an updated list of references on IM-KK theory and related matters 
we again refer the reader to Overduin and Wesson~\cite{Overduin97}.

It has long been known that there are solutions to the Einstein 
field equations which possess causal anomalies in the form of 
closed time-like curves. The famous solution found by 
G\"odel~\cite{Godel49} in 1949 might not be the first but
it certainly is the best known example of a cosmological
model which makes it apparent that the general relativity,
as it is normally formulated, does not exclude the existence of 
closed time-like world lines, despite its Lorentzian character 
which leads to the local validity of the causality principle.
The G\"odel model is a solution of Einstein's field equations
with cosmological constant for incoherent matter (dust) with
rotation. Owing to its striking properties G\"odel's model
has a well-recognized importance and has to a certain extent 
motivated the investigations on rotating cosmological
G\"odel-type models and on causal anomalies in the 
framework of general relativity~\cite{Som68}~--~\cite{Krasinski97} 
and other theories of gravitation~\cite{Vaidya84}~--~\cite{Aman97}.
Among the relevant results on G\"odel-type models in general 
relativity, it is worth mentioning the Bampi-Zordan 
theorem~\cite{Bampi78}, which states that every G\"odel-type 
solution of the 4-D Einstein's equations whose energy-momentum tensor 
is that of a perfect fluid is necessarily isometric to the 
G\"odel model~\cite{Godel49}.

In this work we shall examine the causal anomalies of G\"odel-type 
in the two above-mentioned Kaluza-Klein gravity theories. 
We shall be particularly concerned with the question as to
whether these theories of gravitation allow solutions in which
the causality is violated. To this end, we investigate whether 
these theories admit the 5-D G\"odel-type metrics%
~\cite{ReboucasTeixeira97} as solutions to their field equations.
Perfect fluid and dust G\"odel-type solutions of the STM-KK field 
equations (with and without cosmological constant $\Lambda$) are 
discussed. We also extend the Bampi-Zordan results~\cite{Bampi78} 
to the context of STM-KK theory, i.e., we show that every 
G\"odel-type perfect fluid solution of the STM-KK field equations 
is isometric to the 5-D G\"odel's geometry, which is shown to be the 
only G\"odel-type dust solution of the STM-KK theory.
There emerges from our results that, similarly to the general 
relativity theory, the STM-KK theory%
~\cite{Wesson83}~--~\cite{Wesson84} does not exclude violation
of causality of the G\"odel type. 

In the context of the induced matter theory of gravitation, besides 
examining possible G\"odel-type solutions of the IM-KK field 
equations, we discuss the question as to whether 5-D 
G\"odel-type non-causal geometries discussed in Ref. 
\cite{ReboucasTeixeira97} induce any physically acceptable 
4-D energy-momentum tensor. We show that there is no {\em curved\/} 
G\"odel-type metric which is 
solution to the IM-KK field equations~(\ref{5DfeqsG}), 
making apparent that the non-causal G\"odel-type 5-D curved 
geometries cannot induce any type of matter in 4-D.
Although one cannot yet recommend the 5-D induced matter 
version of general relativity~\cite{Wesson90,WessonLeon92a} 
as an efficient therapy for all sort of causal pathologies that occur 
in general relativity, as it is normally formulated, our results
are certainly an important step towards this conjecture in
that they prove the effectiveness of the induced matter theory as a
therapy for causal anomalies of the 5-D G\"odel-type 
family of metrics. 
 
\vspace{3mm}  
\section{Solutions and Causal Anomalies}
\label{Sols}
\setcounter{equation}{0}
In this section we shall first be concerned with G\"odel-type solutions 
of both Kaluza-Klein theories of gravitation we have discussed in
the Introduction. To this end we write the 5-D G\"odel-type 
line element in the form
\begin{equation} \label{ds2c}
d\hat{s}^{2}=[\,dt+H(r)\, d\Phi\,]^{2} -D^{2}(r)\, d\Phi^{2} -dr^{2} 
                 -dz^{2} - d\psi^2 \;,
\end{equation}
where $H(r)$ and $D(r)$ are arbitrary real functions of $r$, and
the coordinates are such that $x^0=t$ (time), 
$x^{1,2,3}= r, \Phi, z$ (spacelike cylindrical coordinates) and
$x^4 = \psi$ (extra coordinate).

Defining the one-forms $\widehat{\Theta}^A$ according to
\begin{equation} \label{lorpen}
\widehat{\Theta}^{0} = dt + H(r)\,d\Phi\,, \: \quad
\widehat{\Theta}^{1} = dr\,, \: \quad
\widehat{\Theta}^{2} = D(r)\,d\Phi\,, \:\quad
\widehat{\Theta}^{3} = dz \,, \: \quad
\widehat{\Theta}^{4} = d\psi \,,          
\end{equation}
the G\"odel-type line element (\ref{ds2c}) can be written
as 
\begin{equation} \label{ds2f}
d\hat{s}^2 = \widehat{\eta}_{A_{}B_{}} \: 
\widehat{\Theta}^A \,\, \widehat{\Theta}^B = 
(\widehat{\Theta}^0)^2 - (\widehat{\Theta}^1)^2 - 
(\widehat{\Theta}^2)^2 - (\widehat{\Theta}^3)^2 - 
(\widehat{\Theta}^4)^2\,, 
\end{equation}
where here and in what follows capital Latin letters are frame indices 
and run from 0 to 4; they are raised and lowered with Lorentz matrices
$\widehat{\eta}^{AB} = \widehat{\eta}_{AB} = 
                       \mbox{diag} (+1, -1, -1, -1, -1)$,
respectively. 

A straightforward calculation, performed by using the computer 
algebra package {\sc clas\-si}~\cite{Aman87,MacCallumSkea94},
gives the frame components of the Einstein tensor 
$\widehat{G}_{AB}=\widehat{R}_{AB}-\frac{1}{2}\,R\,\widehat{\eta}_{AB}$ 
in the frame (\ref{lorpen}). We find that the nonvanishing components 
are  
\begin{eqnarray}  
\widehat{G}_{00} &=& - \,\frac{D''}{D} + \frac{3}{4}\,
            \left( \frac{H'}{D}\,\right)^2       \label{ein00} \;, \\
\widehat{G}_{02} &=&  \frac{1}{2} \, \left( \frac{H'}{D}\, \right)' \;, 
                                                  \label{ein02} \\  
\widehat{G}_{11} &=& \widehat {G}_{22}\; = \; \frac{1}{4} \, 
                \left( \frac{H'}{D}\, \right)^2\,, \label{ein11} \\
\widehat{G}_{33} &=& \widehat{G}_{44}\; = \; \frac{D''}{D} - \frac{1}{4}\,
            \left( \frac{H'}{D}\,\right)^2         \label{ein33} \;, 
\end{eqnarray}
where the prime denotes derivative with respect to $r$.

The field equations in the STM-KK gravity theory are taken to be 
5-D analogues of the usual 4-D Einstein's field equations%
~\cite{Wesson84,Wesson90}, namely
\begin{equation}  \label{einfrm}
 \widehat{G}_{AB} + \Lambda\, \widehat{\eta}_{AB} = \kappa \;
                             \widehat{T}_{AB}\;,
\end{equation}
where $\Lambda$ and $\kappa$ are the cosmological and the Einstein
constants, respectively. We shall use units such that $\kappa=c=1$, where
$c$ denotes the speed of light.
We shall take the 5-D energy-momentum tensor $\widehat{T}_{AB}$ 
to be that of a comoving perfect fluid~\cite{Wesson84,Gron88b}, viz.,
\begin{equation} \label{Tab}
\widehat{T}_{AB} = \mbox{diag}\,(\,\rho\,, p\,, p\,, p\,, p_4\,)\;,
\end{equation}
wherein  $\rho$ is the matter density,
$p$ is the isotropic pressure and $p_4$ denotes the pressure  
along the extra direction defined by $\widehat{\Theta}^4= d\psi$.

When $\Lambda=0$, the field equations~(\ref{einfrm}) and the 
energy-momentum tensor~(\ref{Tab}) require that $\widehat{G}_{02}=0$, 
which in turn implies
\begin{equation}   \label{G02}
 \frac{H'}{D} = \mbox{const} \equiv - 2\,\omega\;.
\end{equation}
{}From equations~(\ref{ein00})~--~(\ref{ein33}) and (\ref{G02})
one finds that the remaining field equations reduce to
\begin{eqnarray}  
\rho &=& 3\, \omega^2  - \frac{D''}{D}  \;,  \label{dens} \\
 p   &=& \omega^2  \;,    \label{pres} \\
p_4  &= & p \;=\; \frac{D''}{D} - \omega^2  \;. \label{pres4} 
\end{eqnarray}
The equations (\ref{pres}) and (\ref{pres4}) then lead to
\begin{equation}    \label{Deq}
\frac{D''}{D}= \mbox{const} \equiv m^2 \;,
\end{equation}
where for the present solution $m^2 = 2\, \omega^2$.

Now, taking into account the irreducible set of isometrically 
nonequivalent homogeneous G\"odel-type metrics%
~\cite{ReboucasTeixeira97}, the solution to field 
equations~(\ref{G02}) and~(\ref{Deq}) can always be brought into 
the form~(\ref{ds2c}) with

\parbox{14cm}{\begin{eqnarray*} 
D(r)&=&\frac{1}{m}\, \sinh\,(mr)\;,  \\
H(r)&=&\frac{\sqrt{2}}{m}\: [1 - \cosh\,(mr)]\;, 
\end{eqnarray*}}  \hfill
\parbox{1cm}{\begin{eqnarray} \label{sol}  \end{eqnarray}}
where, without loss of generality, we have used $m=\sqrt{2}\,\omega$
for definiteness.

{}From (\ref{dens})~--~(\ref{pres4}) one obtains the equation of 
state
\begin{equation}
p = \rho =  p_4 = \omega^2 \;,  \label{stiff}
\end{equation}
which characterizes a stiff matter type of fluid. In~(\ref{stiff}) the
positivity of the density and of the pressure are manifestly fulfilled.
 
Solutions with nonvanishing cosmological constant $\Lambda$ can
be similarly found. Indeed, here again $\widehat{G}_{02}=0$ 
implies (\ref{G02}), and the remaining STM-KK field equations 
reduce to
\begin{eqnarray}  
\widehat{G}_{00} &=& - \,\frac{D''}{D} +3\,\omega^2 =
                     \rho - \Lambda \;,         \label{G00} \\
\widehat{G}_{11} &=& \widehat {G}_{22}\; = \omega^2 = 
                       p + \Lambda \;,          \label{G11} \\
\widehat{G}_{33} &=& \frac{D''}{D} - \omega^2 =
                       p + \Lambda \;,          \label{G33} \\
\widehat{G}_{44}\;& = &\frac{D''}{D} - \omega^2 =
                       p_4 + \Lambda \;.          \label{G44}
\end{eqnarray}
On the other hand, it is straightforward to show that 
equations~(\ref{G11})~--~(\ref{G33}) lead to (\ref{Deq}), 
which together with (\ref{G02}) gives rise to the 
solution~(\ref{sol}); but now using (\ref{G00}), (\ref{G33}), 
(\ref{G44}) and (\ref{Deq}), instead of the equation of 
state~(\ref{stiff}) one finds
\begin{eqnarray}  
\rho & = & \omega^2 + \Lambda \;, \label{rho} \\
  p  & = & \omega^2 - \Lambda \; =\; p_4  \;, \label{press}
\end{eqnarray}
which does not correspond to a stiff matter type of fluid unless 
$\Lambda = 0$. Here the positivity of the density $\rho$ and 
the pressure $p$ are ensured provided that the cosmological constant
lies in the range $-\omega^2 \leq \Lambda \leq \omega^2 $.
In the case where $0  \leq \Lambda \leq \omega^2 $
we have a linear barotropic equation of state, 
viz.,    
\begin{equation}
p = \gamma \, \rho \;, \qquad     
0 \leq  \gamma \leq 1 \;,
\end{equation}
where $\gamma \equiv (\omega^2-\Lambda)\, / \,(\omega^2+\Lambda)$.  
Clearly the limiting case when $\Lambda =0$ is the previous solution 
given by (\ref{ds2c}), (\ref{sol}) and (\ref{stiff}).

There also exists a G\"odel-type dust solution of STM-KK field 
equations. Indeed, the solution is again given by (\ref{sol}) 
but now, using (\ref{rho}) 
and (\ref{press}) one obtains
\begin{equation}  \label{rhodust}
\rho = 2\, \Lambda = 2\,\omega^2 > 0 \;.
\end{equation}

It should be stressed that the solution~(\ref{sol}) with the density
given by~(\ref{rhodust}) is the only G\"odel-type dust solution of the 
STM-KK field equations. We shall hereafter refer to this solution as
the 5-D G\"odel model.

We shall now show how the above results give rise to the extension
of the Bampi-Zordan theorem~\cite{Bampi78} we have referred to in
the Introduction.
Indeed, the equations (\ref{G02}) and (\ref{Deq}) hold for all the 
above solutions of the STM-KK theory. But for arbitrary constants
$m^2$ and $\omega$, equations (\ref{G02}) and (\ref{Deq}) are 
the necessary and sufficient conditions for a 5-D G\"odel-type manifold 
to be (locally) homogeneous~\cite{ReboucasTeixeira97}.
Therefore, taking also into account the theorem 2 of Ref.\ 
\cite{ReboucasTeixeira97},
the above solutions are (locally) homogeneous and admit a group
of isometry $G_r$ of dimension $r=7$. Moreover, as there is a
fixed relation ($m^2=2\,\omega^2$) between the essential parameters 
$m^2$ and $\omega$ for them all, they characterize (locally) 
just one 5-D STM manifold~\cite{ReboucasTeixeira97},
which is nothing but the 5-D counterpart of the 4-D G\"odel
(dust) model. Thus, all perfect fluid G\"odel-type
solutions of STM-KK field equations~(\ref{einfrm}) are isometric 
to the 5-D G\"odel metric~(\ref{ds2c}) and (\ref{sol}), which is 
the only G\"odel-type dust solution of the STM-KK theory.  

The 5-D G\"odel metric (\ref{ds2c}) with $H$ and $D$ given by
(\ref{sol}) permits violation of causality. Indeed, the line 
element~(\ref{ds2c}) can be rewritten as
\begin{equation} \label{ds2cx} 
d\hat{s}^2=dt^2 +2\,H(r)\, dt\,d\Phi -dr^2 -G(r)\,d\Phi^2 -
                 dz^2 -d\psi^2\,,
\end{equation}
where $G(r)= D^2 - H^2$.  In this form it is clear that the 
existence of closed timelike curves depends on the behaviour
of $G(r)$. If  $G(r) < 0$ within a certain range of $r$ 
($r_1 < r < r_2$, say), G\"odel's circles defined by
$t, z, \psi, r = \mbox{const}\,$ are closed timelike curves.  
Particularly, for the above 5-D G\"odel geometry given by
~(\ref{sol}) and (\ref{ds2cx}), it is easy to find out that
there is a critical radius $r_c$, defined by 
$\sinh(mr_c/2)=1$, such that for all $r_c < r < \infty$ the circles 
$t, z, \psi = \mbox{const}\,$ and $r = \mbox{const} > r_c$ are 
closed timelike curves. Thus, similarly to the general relativity
theory, the STM-KK gravity theory admits solutions which allow the 
violation of causality.

Following the reasoning outlined in the previous paragraph one 
can show~\cite{ReboucasTeixeira97} that there are closed 
timelike curves for all classes of homogeneous 5-D manifolds endowed 
with a G\"odel-type metric~(\ref{ds2c}) [or equivalently~(\ref{ds2cx})].
However, in what follows we shall show that these types of non-causal 
curved manifolds cannot be accomodated in the context of the induced 
matter Kaluza-Klein gravity theory~\cite{Wesson90,WessonLeon92a}. 

It should be noticed from the outset that since $\phi=1$ and the metric 
components of~(\ref{ds2c}) do not depend on the fifth coordinate $\psi$, 
all components of the induced energy momentum-tensor~(\ref{Tinduced}) 
obviously vanish, i.e., the 5-D G\"odel-type 
metrics~(\ref{ds2c}) induce no matter in 4-D. As a matter of fact, 
there is no {\em curved} manifold with 5-D metric~(\ref{ds2c}) 
solution to the field equations~(\ref{5DfeqsG}). To show this, we 
first note that as $\widehat{G}_{02}=0$ then  (\ref{G02}) holds again. 
On the other hand, taking in account~(\ref{ein00})~--~(\ref{ein33}) 
and (\ref{G02}) one easily finds that  the IM-KK field 
equations~(\ref{5DfeqsG}) are fulfilled if and only if
$m^2 = \omega = 0$, which leads to
\begin{equation}
H =  a  \qquad \mbox{and}  \qquad
D =  b\,r + c \;,
\end{equation}
where $a$, $b$ and $c$ are arbitrary real constants. However,
the constants $a$, $b$ and $c$ have no physical meaning, and 
can be taken to be $a = c = 0$ and $b=1$ by a suitable choice of
coordinates. Indeed, if one performs the coordinate transformations 
\begin{eqnarray} 
t   &  = & \bar{t} - \frac{a}{b}\,\, \bar{\Phi}\,, \qquad \quad     
r      =  \bar{r} - \frac{c}{b}\,,   \label{tr}    \\
\Phi & = & \frac{\bar{\Phi}}{b}\,  \,,  \qquad   
z   = \bar{z}  \,,  \qquad  
\psi  =  \bar{\psi}  \,, \label{ppz}
\end{eqnarray}
the line element~(\ref{ds2cx}) becomes
\begin{equation}  \label{ds2flat}
d\hat{s}^2=d\bar{t}^2 -d\bar{r}^2 -\bar{r}^2 \,d\bar{\Phi}^2 -
                 d\bar{z}^2 -d\bar{\psi}^2\;,
\end{equation}
in which we obviously have  $\,G(\bar{r})= \bar{r}^2 >0\,$ for 
$\bar{r} \not=0$. 
The line element (\ref{ds2flat}) corresponds to a 
manifestly flat 5-D manifold, making it clear that the underlying 
manifold can be taken to be the simply connected Euclidean 
manifold $\R^5$, and therefore as $\,G(\bar{r})>0\,$
no closed time-like circles are permitted. 
Furthermore the above results clearly show that the IM-KK theory 
does not admit any {\em curved\/} 5-D G\"odel-type 
metric~(\ref{ds2c}) as solution to its field
equations~(\ref{5DfeqsG}).

However, in a recent work McManus~\cite{McManus94} has shown 
that a one-parameter family of solutions of the  field 
equations~(\ref{5DfeqsG}) previously found by Ponce de Leon%
~\cite{Leon88} (see also Wesson~\cite{Wesson92c,Wesson96a} and
Coley {\em et al.\/}~\cite{Coley95})
was in fact flat in five dimensions, i.e., the 5-D Riemann
tensor vanishes identically. And yet the corresponding 4-D
induced models were shown to be curved. Actually, the 
induced 4-D space-times studied in~\cite{Leon88} constitute 
a perfect fluid family of Friedmann-Robertson-Walker models.
A question which naturally arises here is whether the above 
5-D flat solutions of the IM-KK field equations can
similarly give rise to any 4-D {\em curved\/} space-time.
By using the eqs.\ (5.3) of Ref.~\cite{Reboucas83} 
or simply by using a computer algebra package as, e.g.,  
{\sc clas\-si}~\cite{Aman87,MacCallumSkea94} it is 
straightforward to show that when $m^2=\omega=0$ 
the above 5-D flat geometry [with $m^2$ and $\omega$ defined by 
(\ref{G02}) and (\ref{Deq})] gives rise to nothing but 
the 4-D Minkowski (flat) space-time.

\vspace{3mm}
\section{Concluding Remarks}

In general relativity, the causal structure of 4-D space-time
has locally the same qualitative nature as the flat space-time
of special relativity --- causality holds locally. The global
question, however, is left open and significant
differences can occur. On large scale, the violation of causality
is not excluded. A few familiar space-times make it clear
that general relativity, as it is normally formulated, does not 
phohibit closed time-like curves. The G\"odel model~\cite{Godel49}
is perhaps the best known example of a cosmological solution of
Einstein's equations in which causality may be violated.

In a recent article the main properties of the five-dimensional 
Riemannian manifolds endowed with a 5-D analogue of the 4-D 
G\"odel-type metric were investigated and the main mathematical 
properties discussed~\cite{ReboucasTeixeira97}.
Among several results,  an irreducible set of isometrically 
nonequivalent 5D (locally) homogeneous G\"odel-type 
metrics were exhibited. It was also shown that, apart from the 
degenerated G\"odel-type metric, for which the rotation 
$\Omega^\mu= (0,0,0, \frac{H'}{2D}\,)$ vanishes, in all classes of
homogeneous G\"odel-type geometries there is breakdown 
of causality. As no use of any particular field equations was 
made, these results hold for any 5-D G\"odel-type manifolds 
regardless of the underlying 5-D Kaluza-Klein theory of 
gravitation.

In this work we have investigated the 5-D G\"odel-type geometries
from a more physical perspective. Particularly, we have examined 
the question as to whether the space-time-mass and the induced mass 
theories of gravitation (STM-KK and IM-KK gravity theories%
~\cite{Wesson83}~--~\cite{Wesson90}, \cite{WessonLeon92a}) allow 
solutions that violated causality. We have shown that, similarly
to general relativity, the STM-KK theory permits noncausal 
solutions of G\"odel type. However, the IM-KK gravity theory
is shown to exclude this type of noncausal geometries. 

Two perfect fluid and a dust solutions of the STM-KK gravity theory 
have been found. Actually, we have extended to the context of the STM-KK
theory the Bampi-Zordan results~\cite{Bampi78}, proving that all 
G\"odel-type perfect-fluid solutions of the STM-KK field equations 
are isometric to the unique G\"odel-type dust solution of this 
theory.

Regarding the question as to whether the 5-D G\"odel-type noncausal
geometries~\cite{ReboucasTeixeira97} induce any physically 
acceptable 4-D matter, clearly since $\phi=1$ and the metric components 
of~(\ref{ds2c}) do not depend on the fifth coordinate $\psi$, this
noncausal family of geometries induce no 4-D matter. In agreement with
this fact we have shown that no {\em curved\/}
solution of G\"odel-type is permited in the IM-KK gravity theory. 
Moreover the surviving {\em flat} solution of the IM-KK theory
is shown to give rise to the 4-D Minkowski space-time.

We remark that the 5-D G\"odel-type metric we have
studied is the simplest stationary 5-D class of geometries 
for which $\psi = \mbox{const}$ section is the 4-D G\"odel-type
metric of general relativity. Nevertheless, the 5-D seed 
metric~(\ref{ds2c}) does share the most important
features of the 4-D G\"odel-type counterpart, namely it permits 
violation of causality, admits nonzero rotation $\Omega^\mu$,
and has the most relevant  symmetries ($\partial_t$, 
$\partial_\phi$ and $\partial_z$ are Killing vector fields).
However, as ~(\ref{ds2c}) does not depend on the fifth coordinate
$\psi$ a radiation-like equation of state is an underlying 
assumption of this work regarding the IM-KK theory.
The dependence of the 5-D metric on the extra coordinate gives to
the 5-D IM-KK field equations~(\ref{5DfeqsG}) a rich enough 
structure which permits that matter of a very general type 
can be induced in 4-D.
An exhaustive study of causal anomalies and solutions in a more 
general 5-D geometrical setting has been carried out and we hope 
to publish an account of our results shortly elsewhere. 
We anticipate, however, that if one demands that the 5-D seed 
generalized G\"odel-type metrics share the above-mentioned properties
of the 4-D G\"odel-type counterpart, it can be shown that 
the IM-KK theory also excludes a large class of noncausal generalized
G\"odel-type stationary geometries with two arbitrary functions 
$F(\psi)$ and $G(\psi)$ of the fifth coordinate $\psi$, besides,
obviously, the functions $H(r)$ and $D(r)$. 

Although one cannot yet recommend the 5-D induced matter 
Wesson's version of general relativity~\cite{Wesson90,WessonLeon92a} 
as an efficient treatment for all sort of causal anomalies in general 
relativity, our results constitute an important step in this direction 
inasmuch as they make apparent the effectiveness of the induced matter 
theory as a therapy for causal anomalies of the 5-D 
G\"odel-type families of metrics discussed in this work.

\vspace{3mm}
{\raggedright
\section*{Acknowledgement} } \label{acknowl}  
\setcounter{equation}{0}
The authors gratefully acknowledge financial assistance from CNPq.

\end{document}